Title: Kolmogorov turbulence in a random-force-driven Burgers equation:
anomalous scaling and probability density functions

\documentstyle [12pt]{article}
\newfont{\feff}{cmti10}
\begin{document}

\input{psfig.sty}

\title{\bf Kolmogorov turbulence in a random-force-driven Burgers
equation: anomalous scaling and probability density functions}

\author{\\{Alexei Chekhlov and Victor Yakhot} \\ \\{\em Program in
Applied and Computational Mathematics,} \\ {\em Princeton
University,}\\ {\em Princeton, New Jersey 08544, USA}}

\maketitle

\psfull

\newcommand{\pdf}{{\cal P}}

\begin{abstract}
High-resolution numerical experiments, described in this work, show
that velocity fluctuations governed by the one-dimensional Burgers
equation driven by a white-in-time random noise with the spectrum
$\overline{\left|f(k)\right|^{2}}\propto k^{-1}$ exhibit a biscaling
behavior: All moments of velocity differences $S_{n\leq
3}(r)=\overline{\left|u(x+r)-u(x)\right|^n}\equiv\overline{\left|\Delta
u\right|^n}\propto r^{n/3}$, while $S_{n>3}(r)\propto r^{\xi_n}$ with
$\xi_n\approx 1$ for real $n>0$ (Chekhlov and Yakhot, Phys. Rev. E
{\bf 51}, R2739, 1995). The probability density function, which is
dominated by coherent shocks in the interval $\Delta u<0$, is
$\pdf(\Delta u,r)\propto(\Delta u)^{-q}$ with $q\approx 4$. A
phenomenological theory describing the experimental findings is
presented.
\end{abstract}
PACS number(s): 47.27.Gs.

\vspace{0.3in}

Our recent study \cite{burg} of the one-dimensional Burgers equation
\begin{eqnarray}
\frac{\partial u}{\partial t}+u\,\frac{\partial u}{\partial
x}=f+\nu\,\frac{\partial^2 u}{\partial x^2}
\end{eqnarray}
\noindent driven by a white-in-time random force defined by the
correlation function
\begin{eqnarray}
\overline{f(k,t)f(k',t')}\propto D(k)\,\delta(k+k')\,\delta(t-t')
\end{eqnarray}
\noindent with $D(k)=\epsilon_0\,k^{-1}$ and $\epsilon_0=O(1)$ was
motivated by an interest in dynamical
processes which involve an interplay between chaotic and coherent
phenomena. It has been shown that the velocity field $u(x,t)$ consists
of random-in-time and random-in-space fluctuations superimposed on the
relatively strong and long-living shocks. Numerical simulations
yielded the energy spectrum
$E(k)\propto\overline{\left|u(k)\right|^2}\propto k^{-x}$ with $x=5/3
\pm 0.02$, characteristic of Kolmogorov turbulence \cite{Kolmogorov}
and the Eulerian correlation function
$C(k,\omega)=\overline{\left|u(k,\omega)\right|^2}\propto
k^{-7/3}\,\Phi(\omega/k^z)$ with the dynamic exponent $z=2/3$. This
result shows that in this system the kinematic transport of the
small-scale velocity fluctuations by the large-scale structures is
very weak. Investigation of the velocity structure functions
$S_n(r)=\overline{\left[u(x+r)-u(x)\right]^n}\equiv\overline{(\Delta
u)^n}$ with integer $n$ revealed strong deviations from the Kolmogorov
picture of turbulence: all moments $S_{n>3}(r)\propto r^{\xi_{n}}$
with $\xi_{n}\approx 1$, characteristic of strong shocks. Thus, the
system governed by $(1)-(2)$ shows both ``normal'' (Kolmogorov) and
anomalous scalings with the latter dominated by the coherent
structures (shocks). In this work we are interested in the details of
the probability density functions (PDF's) characterizing the
fluctuations generated by $(1)-(2)$ and in the role the structures
play in the determination of the PDF's shape. The PDF
$\pdf(\Delta u,r)$ is defined such that $\pdf(X,r)\,dX$ is the
probability of finding a velocity difference $\Delta u=u(x+r)-u(x)$
within the interval $(X,X+dX)$ for infinitesimally small $dX$. A
spectral code with $12288$ Fourier modes was used in the numerical
experiment.  Equation $(1)$ with a hyperviscous (instead of viscous)
dissipation term was solved. The details of the numerical procedure
are reported in
\cite{burg}.

The most prominent feature of Burgers equation is a tendency to create
shocks and, consequently, to increase the negative velocity
differences $\Delta u<0$ and to decrease the positive ones $\Delta
u>0$ \cite{Burgers}. Thus, strong asymmetry of the curve $\pdf(\Delta
u,r)$ is expected. The two-point PDF $\pdf(\Delta u,r)$ was measured
for a set of separations $r$ covering a variety of scales in the
system in the following way. The range of variation of the velocity
difference, $-5<\Delta u/u_{rms}< 5$, was divided into $10^{4}$
bins. The data were collected during a time longer than $10$ largest
eddy turnover times (corresponding to $O(10^{7})$ time-steps) and were
distributed among the appropriate bins to generate a
histogram. Fig. $1$ presents $\pdf(\Delta u,r)$ for the inertial range
separations $r/dx=200$, $250$, $300$, $350$, $400$, where $dx=L/12288$
is the mesh size and $L=2\pi$ is the system size. It follows from $(1)-(2)$
that: $\overline{(\Delta u)^3}\propto \epsilon_0\,r\,log~rk_{d}$ and
that is why this PDF has a shifted maximum, approximately at
$\phi=(\Delta u)/R^{1/3}\approx 0.5$. Here the function $R(r)$ defined
as $R(r)\equiv\int\overline{\left[f(x+y)-f(x)\right]^2}\,dy$, was also
directly measured. It is fully force-dependent and in a system with
viscosity it may be analytically calculated for the inertial-range
values of separation $r$, giving $R(r)\propto\overline{(\Delta u)^3}
\propto\epsilon_0\,r\,\log(r\,k_d)$, where $k_d\rightarrow+\infty$ is
a dissipative cutoff wavenumber. Technically speaking, the system
considered in this work does not have a real ``inertial range'' since
the mean dissipation rate $\overline{\epsilon}=
O(\epsilon_0\,log(L\,k_{d}))$ depends on the ultra-violet cut-off
$k_d$. This dependence, however, is weak and in what follows we take
$k_{d}=O((\epsilon_0/\nu^3)^{1/4})$.

The tail of the PDF $\pdf(\Delta u,r)$ for $\Delta u <0$ is shown on
Fig. $2$ for various magnitudes of the displacement $r$ in the
universal range. One may observe from this figure that the PDF for
$\Delta u< 0$ may be well approximated as:
\begin{eqnarray}
\pdf(\Delta u,r)\propto (\Delta u)^{-q},
\end{eqnarray}
with $q\approx 4$. We have also found that $\pdf(\Delta u,r)$ for $\Delta
u> 0$ is well fitted by the exponential:
\begin{eqnarray}
\pdf(\Delta u,r)\propto e^{-\alpha\,\frac{(\Delta u)^3}{r}}, \
\end{eqnarray}
with the constant $\alpha$ to be determined from the theory. The
dynamic argument leading to $(3)$ will be presented below. The results
shown on Figs. $1$, $2$ are highly nontrivial because the observed
algebraic decay of the PDF $\pdf(\Delta u,r)$ as $\Delta u/(\Delta
u)_{rms}\rightarrow -\infty$ leads to the divergence of the moments
$S_n(r)$ for $n>3$ for the inviscid case. However, as we also
observed, the single point PDF $\pdf(u)$ is a very rapidly decreasing
function which is close to the Gaussian and that is why the occurrence
of shocks with an amplitude $\Delta u>U_0\approx
(\overline{\epsilon}\,L\,log(L\,k_d))^{1/3}$ is highly improbable and
one can expect the PDF $\pdf(\Delta u,r)$ to decrease sharply for
$\Delta u<-U_0$. This is sufficient for the existence of all moments
$S_n(r)$. A full analytical theory leading to an expression for
$\pdf(\Delta u,r)$, which unifies both asymptotics $(3)$ and $(4)$
will be published elsewhere \cite{Polyakov}.

To develop a phenomenological theory we assume that the flow can be
represented as a superposition of coherent and random components.  The
coherent contribution is visualized as a ``gas of shocks'' and a
single structure (shock) can be approximated by the exact
$\tanh$-solution of the unforced problem \cite{Burgers}. In
particular, let us assume that solution for the normal (not the
hyper-) viscosity case has the form
\begin{eqnarray}
u(x,t)=-\sum_{i=0}^NU_i\,\tanh\left[\frac{(x-a_i)\,U_i}{2\nu}\right]+\phi(x,t).
\end{eqnarray}
The first contribution to the right side of $(5)$ describes the slowly
varying coherent ``gas of shocks'', whereas the second represents the
effects unaccounted for by the first term. Here $a_i$ and $U_i$ denote
the coordinates of the centers of the shocks and the shock amplitudes
respectively. The physical picture behind this representation is the
following: the forcing produces the low energy excitations which
coagulate into ever stronger well separated shocks due to the
non-linear interactions.  It will be clear below that the detailed
shape of the shock assumed in $(5)$ is unimportant. The most essential
feature of the $\tanh$-solution $(5)$ is that the shock width
$l_i\approx\nu/U_i$, which means that the stronger the shock, the more
narrow it is.

Statistics of the dissipation rate fluctuations were investigated in
detail in Ref. $1$. It has been shown that the energy dissipation
takes place mainly ($\approx 99\%$) inside the well separated strong
shocks. Thus, it follows from $(5)$ that the dissipation rate in
interval of length $r$ is
\begin{eqnarray}
\epsilon_r=\frac{1}{4\,r\,\nu}\,\int_{x=0}^r dx\,
\sum_{i,j=0}^N\frac{U_i^2\,U_j^2}{\cosh^2~Y_i\,\cosh^2~Y_j},
\end{eqnarray}
where we denote $Y_i=(x-a_i)\,U_i/(2\nu)$. The principle contribution
to the sum comes from the strong and narrow shocks, and, therefore, we
can neglect the nondiagonal terms with $i\ne j$. Taking the integral
for inertial-range values of $r$ we have
\begin{eqnarray}
\epsilon_r\propto\sum_{i=0}^N\frac{U_i^3}{r}\approx\frac{\overline{U^{3}}}{r}.
\end{eqnarray}
On the other hand, it can be directly shown from $(1)-(2)$ that
\begin{eqnarray}
\epsilon_r=\epsilon_0\,ln\left(\frac{L\,U_0}{\nu}\right).
\end{eqnarray}

\noindent Introducing the PDF $\pdf(U,r)$ to
find a shock with amplitude $U$ in the interval of the length $r$ we
obtain from the last two relations
\begin{eqnarray}
\int_{\frac{\nu}{L}}^{U_0}U^3\,\pdf(U,r)\,dU
\propto\epsilon_0r\,ln\left(\frac{L\,U_0}{\nu}\right),
\end{eqnarray}
from which we readily establish the form of $\pdf(U,r)$
\begin{eqnarray}
\pdf(U,r)\propto\frac{\epsilon_0\,r}{U^4}.
\end{eqnarray}
Since $\pdf(U,r)=\pdf(U)\,r/L$, the relation $(10)$ establishes the
shape of PDF $\pdf(U)\propto U^{-4}$ to find a shock of the
amplitude $U$. Note that $r/L$ is the probability to find a shock
center within the interval of the length $r$ which is in turn placed
in the larger interval of the length $L$. The low integration limit
$U\approx\nu/L$, corresponds to the amplitude of the
``weakest structure'', contributing to $\epsilon_r$. Formula $(10)$ is
a consequence of relations $(7)$ and $(8)$, and is valid in the
logarithmic case when the forcing function is defined by $(2)$. It is
only in this case that we can establish the form of the PDF.

Thus, according to the data presented in Fig. $1$ and the theoretical
considerations developed above, the PDF of velocity differences can be
represented as:
\begin{eqnarray}
\pdf(\Delta u,r)=a\,R^{-\frac{1}{3}}\,F(\frac{\Delta u}{R^{\frac{1}{3}}})
\end{eqnarray}
\noindent
in the interval $O(-1)<x\equiv\Delta u/R^{1/3}<\infty$, where function
$R(r)$ is defined above, $a$ is a numerical constant and $F(x)$ is a
scaling function (see, \cite{Parisi}) which is assumed to go rapidly
to zero when $|x|$ is large.  In the interval $x\ll -1$, and $|\Delta
u|< O(U_0)$, where the PDF is dominated by the well-separated shocks,
we have:
\begin{eqnarray}
\pdf(\Delta u,r)=b\,\frac{\epsilon_0\,r}{(\Delta u)^4},
\end{eqnarray}
\noindent
where $b$  is a constant. When $\Delta u\ll -U_0$, the PDF is
a rapidly decreasing function of $\Delta u/U_0$. The moments of
velocity difference are evaluated readily with the result:
\begin{eqnarray}
S_n(r)=\int \left(\Delta u\right)^n\,\pdf(\Delta u,r)\,d\Delta u=
b_{n}\,r\,\epsilon_0\,\frac{U^{n-3}_0-(\epsilon_{0}R)^{\frac{n}{3}-1}}{n-3}
+ B_n(\epsilon_0\,R)^{\frac{n}{3}},
\end{eqnarray}
\noindent
where the amplitudes $B_n$ depend on the shape of the scaling function
$F(x)$. The constants $b_n\propto (-1)^{n}$ for integer $n$ and for
the noninteger values of $n$ the structure functions
$S_n=\overline{|\Delta u|^n}$ so that relation $(13)$ should include
the absolute value of the first term in the right side. It follows
from $(11)-(13)$ that all moments $S_n(r)$ with $n>3$ are completely
determined by the upper cut-off in $(13)$
\begin{eqnarray}
S_n(r)\propto\epsilon_0\,r\,\frac{U_0^{n-3}}{n-3},
\end{eqnarray}
which is in excellent quantitative agreement with \cite{burg}, whereas
for $0\le n\le 3$:
\begin{eqnarray}
S_n(r)\propto \left(\epsilon_0\,R\right)^{\frac{n}{3}},
\end{eqnarray}
as in the Kolmogorov theory of turbulence \cite{Kolmogorov}. Thus, the
anomalous scaling of the velocity structure functions $S_n(r)$ appears
only for $n>3$. It should be stressed that, in accord with $(13)$, in
the logarithmic case considered in this work the contribution from
the shocks to the moments $S_{n<3}$ is smaller than the one from the
scaling component of the PDF only by factor $1/\log(r\,k_d)$ which makes
the experimental investigation of the details of the crossover very
difficult.

The prediction $(13)$ has been tested in \cite{burg}. It has been
shown that $S_{2n}(r)\propto r^{\xi_{2n}}$ with $\xi_{2n}\approx 0.91$
for $n>2$, indicating that these correlation functions are dominated
by coherent shocks. The results of the measurements of the structure
functions $S_n(r)$ with $n=1/3,2/3,\ldots,6/3$, presented on Fig. $3$,
are in good agreement with the scaling law $(15)$. The general
structure of the moments of velocity differences given by expression
$(13)-(15)$ is similar to the outcome of the recent theories of the
random- force- driven Burgers equation by Polyakov \cite{Polyakov}
(1d) and Bouchaud \cite{Parisi} ($d\rightarrow\infty$) cases. The
Polyakov theory confirmed our qualitative argument leading to the
algebraic decrease of the PDF in the interval $\Delta u<0$.

Fig. $4$ presents the PDF of the shock amplitudes. The problem of the
shock location was solved in the following simple but reliable way. At
each spatial point $x$ the local gradient of the solution $u(x)$ was
measured. Then, if $u'(x)\ge 0$, it was assumed that this point $x$ is
outside of a shock, otherwise $x$ lies inside of a shock. Once inside
a shock, one can march in $x$ until the gradient becomes zero, and
thus the boundaries of the shock may be located, and so forth.  Note
that the shock amplitude obtained in this way has been corrected to
exclude the Gibbs phenomenon typical in spectral approximations of
discontinuous functions. To reduce the statistical noise in $\pdf(U)$
in Fig. $4$, a simple smoothing procedure was applied: $\pdf(U)$ was
averaged over eight surrounding points. The result presented in
Fig. $4$ demonstrates that
\begin{eqnarray}
\pdf(U)\propto U^{-4}
\end{eqnarray}
is observed for all $\left|U/U_{rms}\right|>0.5$. The fact that
$\pdf(\Delta u)\approx \pdf(U)$ when $\Delta u <0$ tells us that in
this range $\pdf(\Delta u)$ is dominated by the well-separated
shocks. This confirms the main assumption of the phenomenological
theory presented above. It follows from Figs. $1-4$ and relation
$(12)$ that the anomaly in the high-order moments results only from
the slow (algebraic) decrease of the PDF in the interval $\Delta
u<0$. As was pointed out above, in this case one expects a cut-off at
some $\Delta u\approx U_0$.

We have also investigated the problem $(1),$ $(2)$ driven by the
white-in-time random forces with $D(k)\neq 0$ only for $k<5$ and
$D(k)\propto k^{-3/2}$ \cite{burg2}. The outcome of the simulations in
both cases revealed the algebraically decreasing $\pdf(\Delta
u,r)\propto r/\left|\Delta u\right|^q$ for $\Delta u/(\Delta
u)_{rms}\ll -1$, with the exponent $q$, related to the functional form
of $D(k)$.  The former case of the large-scale driven Burgers equation
was investigated in a recent paper by Bouchaud et. al. \cite{Parisi}
using a replica tric in the limit of the space dimensionality
$d\rightarrow \infty$. Although the scaling of the moments of velocity
differences, obtained in Ref. $5$ is the same as the one observed in
our simulations, the shape of the PDF in the 1d-case, numerically
found by us, differs dramatically from $\pdf(\Delta u,r)=(1-r)\,
\delta(\Delta u-r) +\beta\,\,F(\Delta u/U_{0})$, derived in Ref. $5$. Here
$F(x)$
is a scaling function and $\beta$ is a number. This means that the
physical mechanisms, responsible for the anomalous scaling in the one
and multi-dimensional systems are different and understanding of the
transition between the two behaviors is an extremely interesting
challenge. The detailed theoretical and numerical investigations of
the different cases of forcing functions will be published elsewhere
\cite{burg2}.

This work was supported by grants from ARPA and AFOSR. We would like
to thank R. Kraichnan for pointing out some inconsistencies in the
preliminary versions of the text. Stimulating discussions with A. Migdal,
S. Orszag, A. Polyakov and Ya. Sinai are gratefully acknowledged.


\begin{figure}[h]
\centerline{\psfig{file=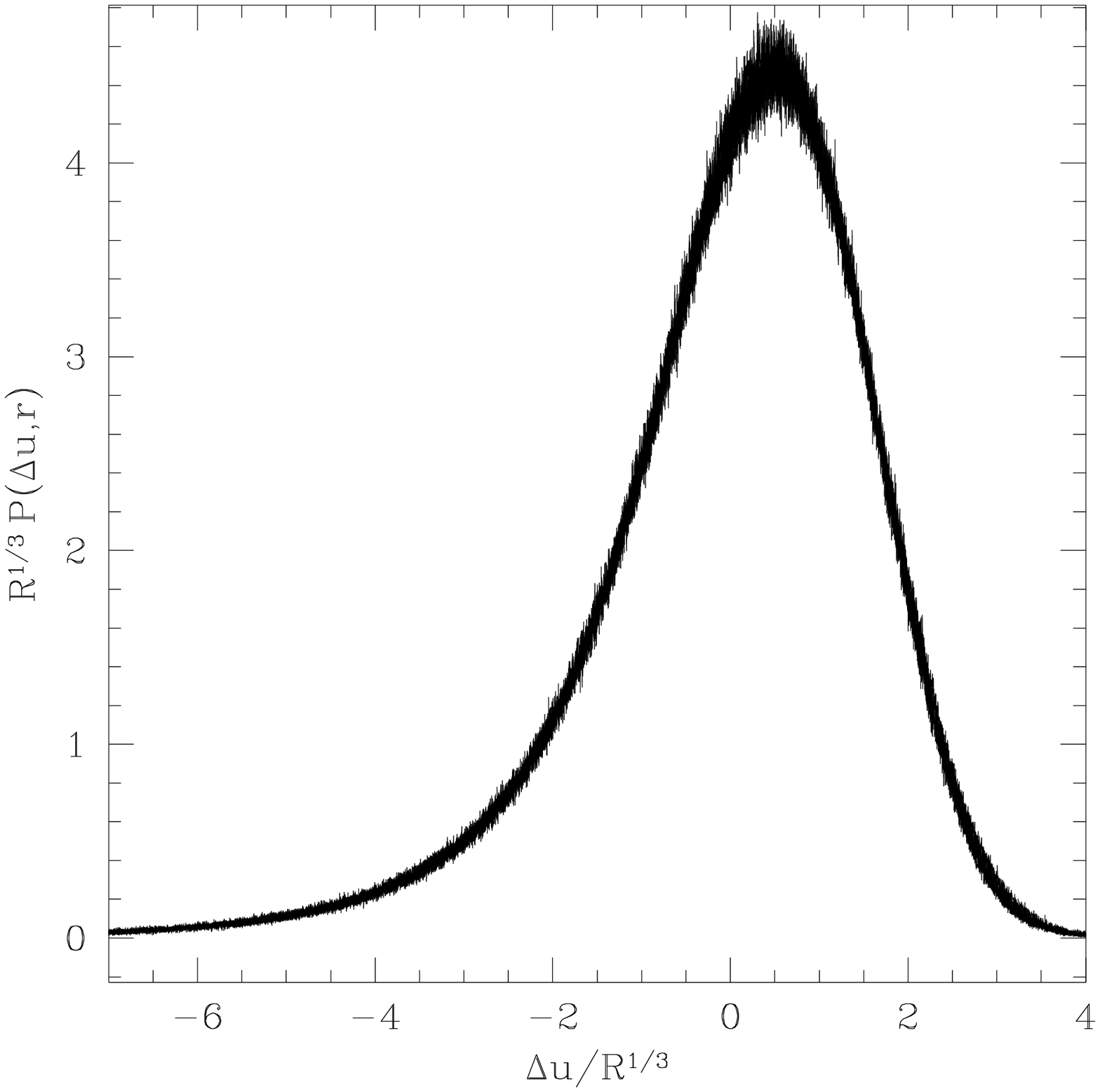,width=6.0in,height=6.0in}}
\caption{Normalized two-point PDF $F(\Delta
u/R^{1/3})=R^{1/3}\,\pdf(\Delta u,r)$ for separations $r/dx=200$,
$250$, $300$, $350$, $400$ within the universal range. The collapse of
various curves supports the choice of the scaling variable
$\phi=(\Delta u)/R^{1/3}$, where function $R(r)$ is defined in the
text.}
\end{figure}


\begin{figure}[h]
\centerline{\psfig{file=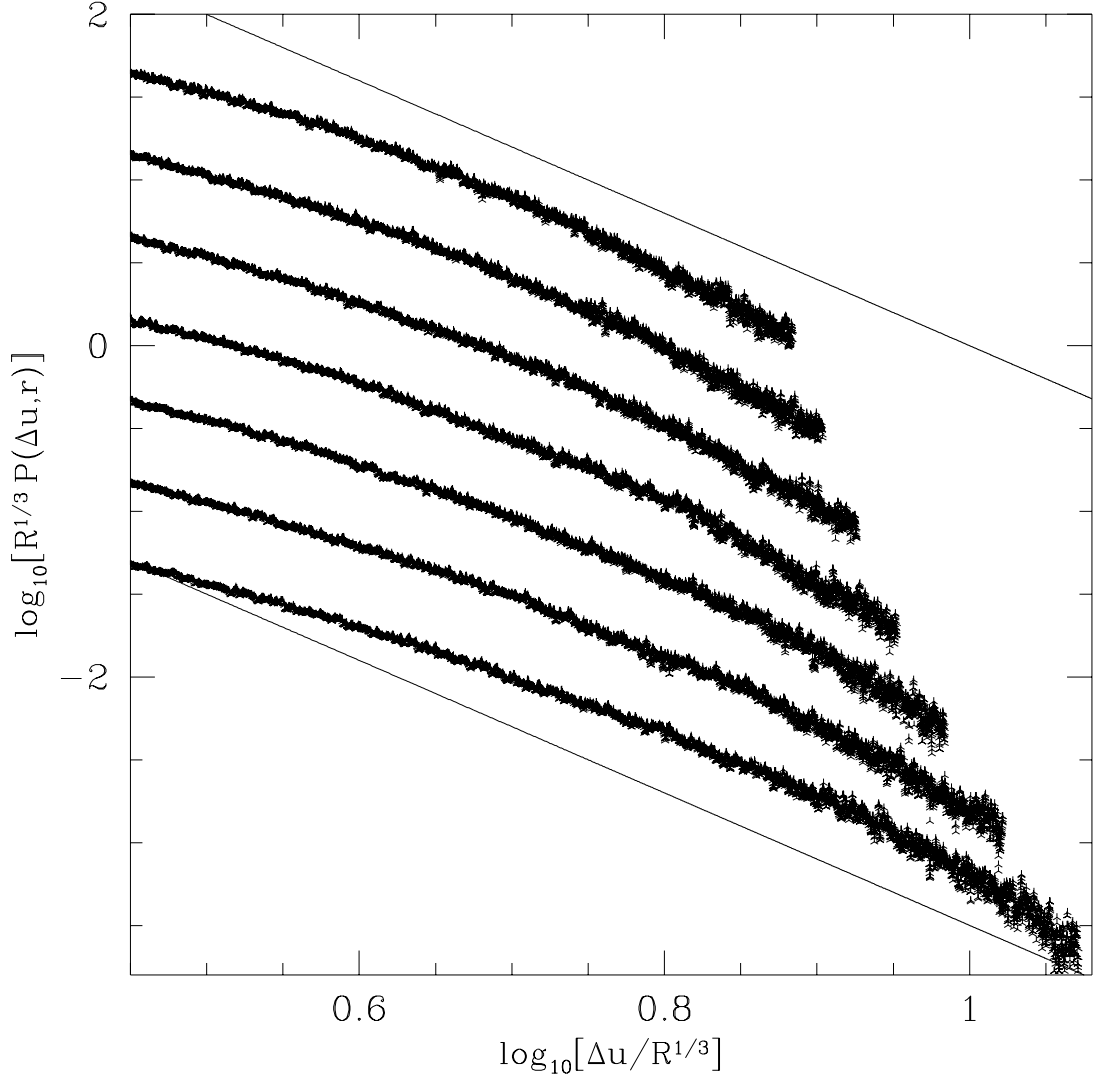,width=6.0in,height=6.0in}}
\caption{The tail of the two-point PDF $\pdf(\Delta u,r)$ (points) for
separations $r/dx=150,200,250,300,350,400,450$ within the universal
range, plotted on a logarithmic-logarithmic scale. The slope of the
solid lines is equal to $-4$. The graphs for different values of $r$
are arbitrarily shifted along the vertical axis for clarity. }
\end{figure}


\begin{figure}[tbp]
\centerline{\psfig{file=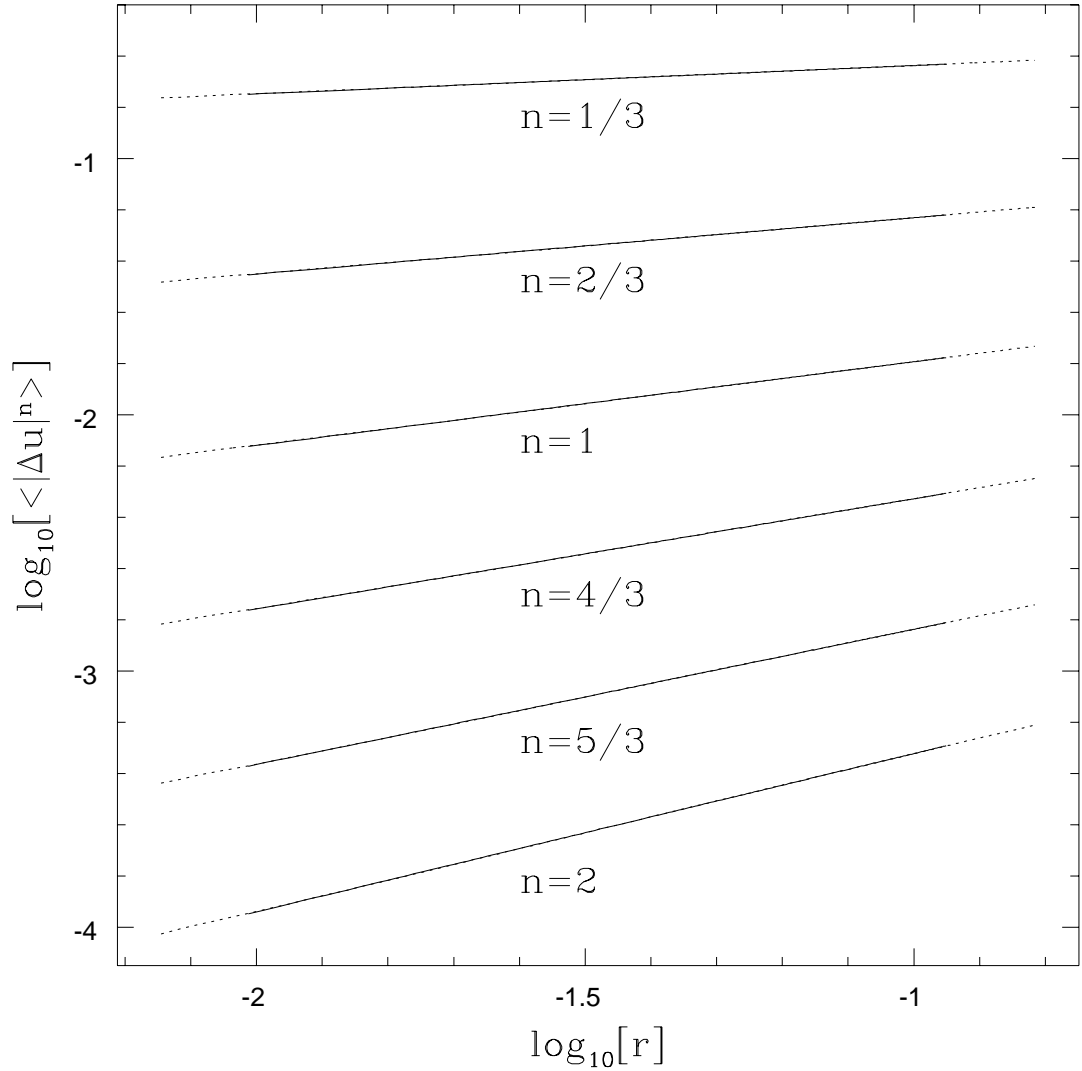,width=6.0in,height=6.0in}}
\caption{Velocity structure functions
$\overline{\left|\Delta u\right|^n}$ for noninteger values
$n=1/3,2/3,\ldots,6/3$ (dotted curves). Slopes of the linear least
square fits (solid lines) from top to bottom respectively are:
$0.111$, $0.222$, $0.330$, $0.433$, $0.531$, $0.620$. }
\end{figure}


\begin{figure}[h]
\centerline{\psfig{file=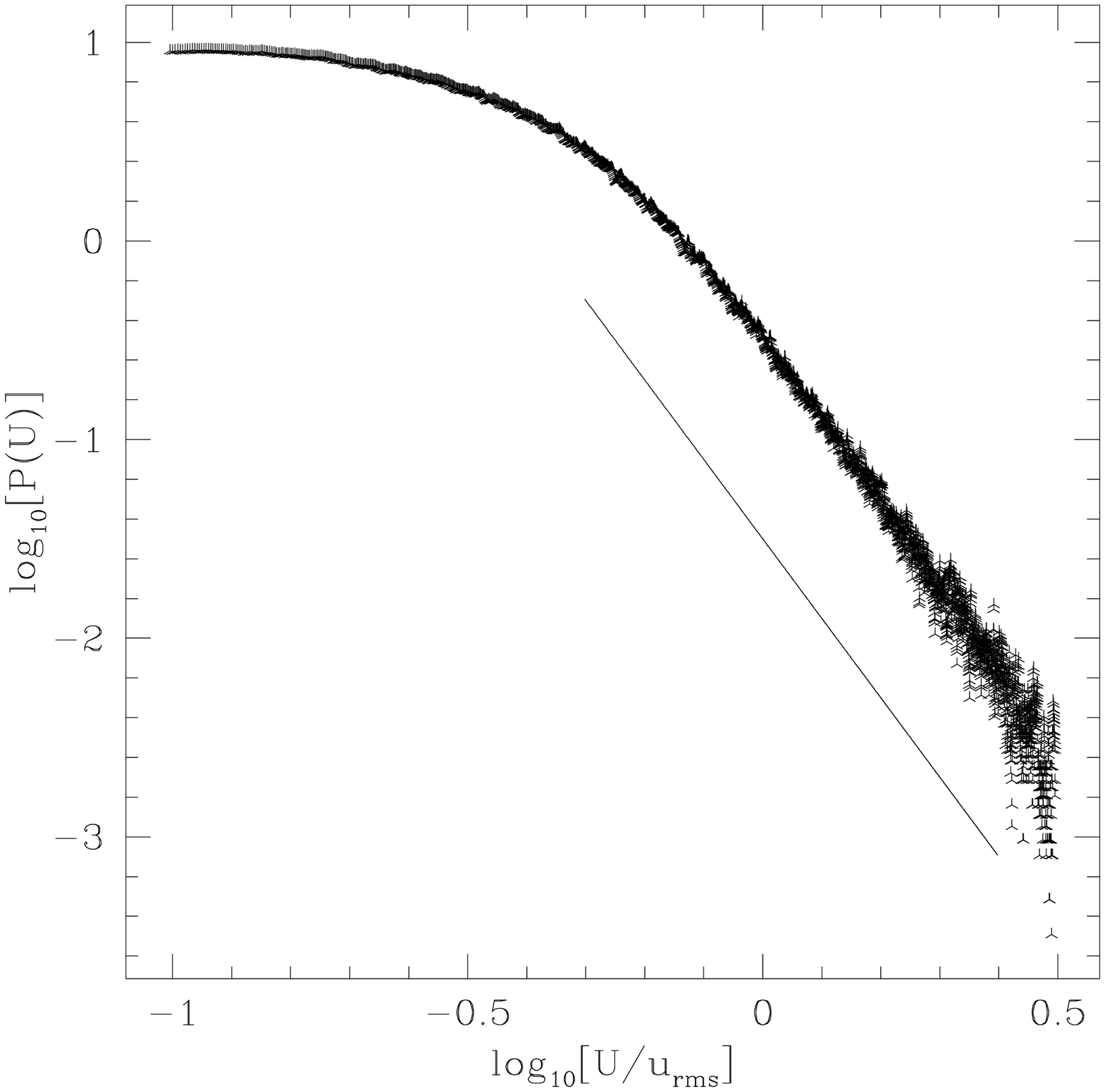,width=6.0in,height=6.0in}}
\caption{PDF of shock amplitudes, $\pdf(U)$ on a
logarithmic-logarithmic scale (points). The slope of the solid line is
equal to $-4$. }
\end{figure}


\begin{thebibliography}{99}

\bibitem{burg} A.~Chekhlov and V.~Yakhot, Phys. Rev. E, {\bf 51},
R2739 (1995).

\bibitem{Kolmogorov} A.~Kolmogoroff, C.~R. (Doklady) de~l'Acad. des
Sci. de l'URSS, {\bf 30}, 301 (1941); {\it ibid} {\bf 32}, 16 (1941).

\bibitem{Burgers} J.~M.~Burgers, {\em The Nonlinear Diffusion Equation.
Asymptotic Solutions and Statistical Problems.} (Reidel, Dordrecht,
1974); J.~Krug and H.~Spohn, in {\it Solids Far From Equilibrium:
Growth, Morphology and Defects}, edited by C.~Godriche (Cambridge
University Press, Cambridge, England, 1992).

\bibitem{Polyakov} A.~Polyakov, unpublished (1995).

\bibitem{Parisi} J. P. Bouchaud, M. M\'ezard, and G. Parisi, Phys.
Rev. E, in press (1995).

\bibitem{burg2} V. Yakhot and A. Chekhlov, unpublished (1995).

\end{thebibliography}
\end{document}